\documentclass[aps,prl,twocolumn,showpacs,amsmath,amssymb,superscriptaddress]{revtex4-1}
\usepackage{graphicx}
\usepackage{dcolumn}
\usepackage{epstopdf}
\usepackage{bm}
\usepackage{color}

\begin{document}
\title{Exchange interactions and frustrated magnetism in single-side hydrogenated and fluorinated graphene}

\author{A.~N. Rudenko}
\email[]{rudenko@tu-harburg.de}
\author{F.~J. Keil}
\affiliation{Institute of Chemical Reaction Engineering, Hamburg University of Technology, Eissendorfer Strasse 38, D-21073 Hamburg, Germany}
\author{M.~I. Katsnelson}
\affiliation{Institute for Molecules and Materials, Radboud University Nijmegen, Heijendaalseweg 135, 6525 AJ Nijmegen, The Netherlands}
\author{A.~I. Lichtenstein}
\affiliation{Institute of Theoretical Physics, University of Hamburg,
Jungiusstrasse 9, D-20355 Hamburg, Germany}
\date{\today}

\begin{abstract}
Magnetism in single-side hydrogenated (C$_2$H) and fluorinated (C$_2$F) graphene is analyzed in terms of the Heisenberg model with parameters determined
from first principles. We predict a frustrated ground state for both systems, which means the instability of collinear spin structures and
sheds light on the absence of a conventional magnetic ordering in defective graphene demonstrated in recent experiments. Moreover, our findings suggest
a highly correlated magnetic behavior at low temperatures offering the possibility of a spin-liquid state.
\end{abstract}

\maketitle


Magnetism of $sp$ materials is an exotic phenomenon of a fundamental physical relevance and potential practical importance. In particular, one can expect much higher Curie temperatures for $sp$ magnets than for conventional magnetic semiconductors \cite{EK2006}. A discovery of graphene has revived an interest in this subject \cite{Yazyev2010,Katsbook}.  Recent studies provided persuasive experimental evidence for the existence of magnetism
in graphene doped by hydrogen and fluorine atoms \cite{Hong,McCreary,Nair2012}. Although no magnetic ordering has been detected in defective graphene even at liquid helium
temperatures \cite{Nair2012,Sepioni}, theoretical studies show that an ordered decoration of graphene with a composition of C:H(F)=2:1 (occasionally referred to as graphone) is unstable toward the formation of (anti-)ferromagnetism \cite{Zhou,Boukhvalov2010}, offering perspectives of applications \cite{L-Li}.

In contrast to single defects in graphene, whose magnetic properties are theoretically well understood \cite{Yazyev2007,Kogan}, the magnetism in C$_2$H and C$_2$F lacks a consistent description.
Although previous density functional (DF) studies point to the possibility of magnetic ordering in C$_2$H and
C$_2$F \cite{Zhou,Boukhvalov2010,L-Li}, its origin and stability have not yet been analyzed in details. Moreover, the prediction of the magnetic ground
state cannot be conclusive within the DF scheme due to symmetry constraints imposed by periodic boundary conditions.
On the other hand, the possibility of a description in terms of spin models is not obvious due to the absence of a unique definition of local magnetic moments in $sp$ systems, which makes the
parametrization of a model particularly challenging.

In this Letter, we construct an appropriate Heisenberg model to describe magnetic properties of C$_2$H and C$_2$F, and determine its parameters
from first principles.
For both systems, we find a frustrated ground state, which contradicts previous theoretical predictions.
We show that the possibility of spontaneous magnetization in C$_2$H and C$_2$F is very unlikely, instead, an unconventional magnetic behavior is expected at low temperatures,
such as non-collinear ordering or a spin liquid state.

\begin{figure}[bp]
\includegraphics[width=0.43\textwidth, angle=0]{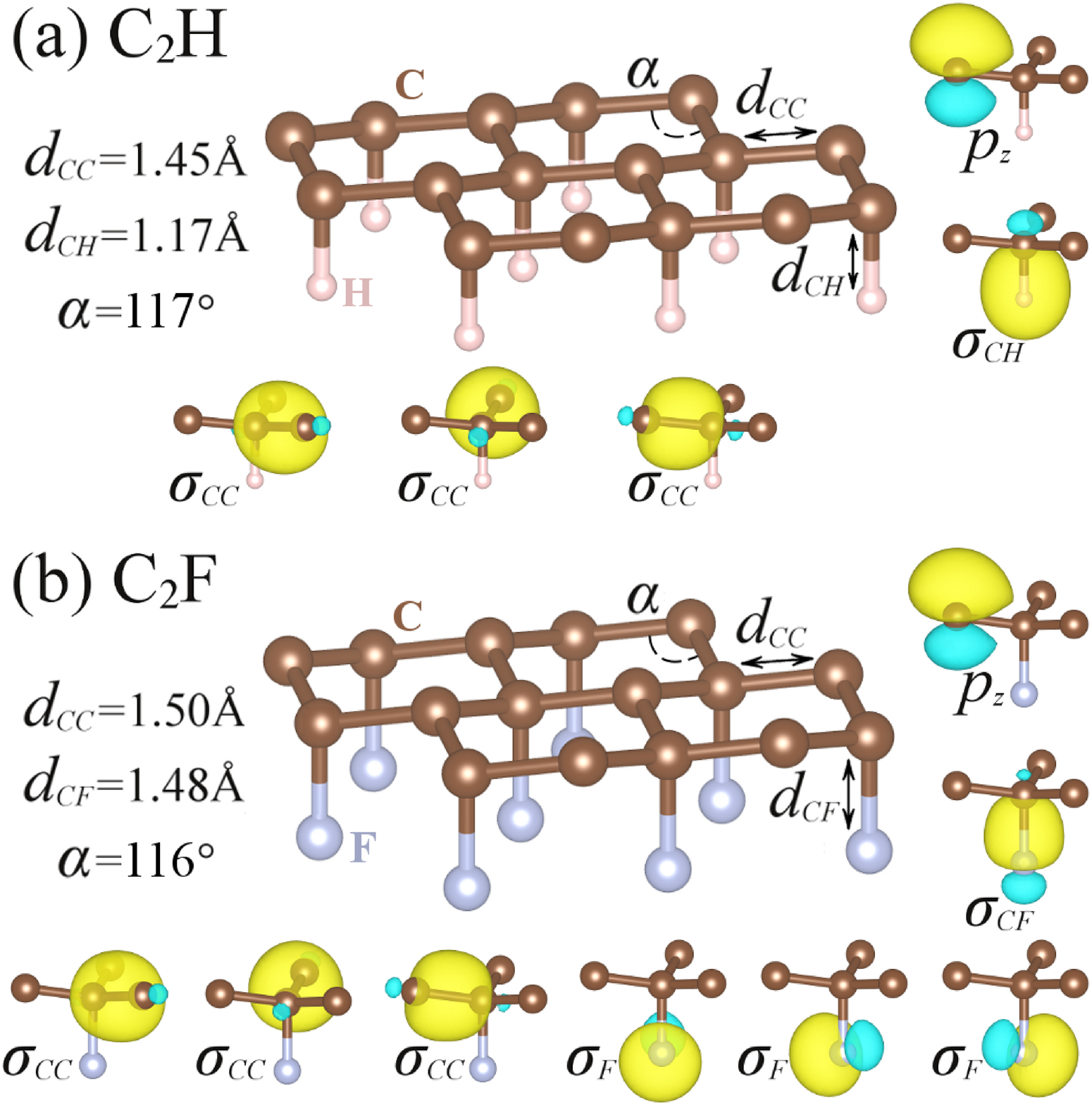}
\caption{(Color online) Schematic representation of the atomic structure and Maximally-Localized Wannier Functions for C$_2$H and C$_2$F.}
\label{wfs}
\end{figure}

We first analyze the electronic structure at the DF level and construct a tight-binding Hamiltonian by taking advantage of the Maximally-Localized
Wannier Functions (MLWF) \cite{Marzari}.
DF calculations were performed within the plane-wave pseudopotential method as implemented in
the {\sc quantum-espresso} simulation package \cite{espresso}. Exchange-correlation effects were taken into account by using the
standard GGA-PBE functional \cite{pbe}.
To achieve high accuracy, we employed an energy cutoff of 50 Ry for the plane-wave basis and 400 Ry for the charge density, as well as
a (64$\times$64) {\bf k}-point mesh. The surface layers were separated by a vacuum region of 40~\AA~and fully relaxed.
To construct a localized basis for the electronic Hamiltonian, we used the {\sc wannier90} code \cite{wannier90}.

Fully relaxed structures of C$_2$H and C$_2$F as well the MLWF are shown in Fig.~\ref{wfs}.
All orbitals are well localized providing an intuitive picture of the chemical bonding.
In each case, the MLWF basis includes three bond-centered orbitals,
corresponding to the $sp^2$-like hybridization of carbon atoms ($\sigma_{CC}$), one bond-centered $sp$-like orbital localized between carbon and hydrogen/fluorine atoms
($\sigma_{CH}/\sigma_{CF}$), and one $p_z$-like orbital centered at the nonbonded carbon atom. In the case of C$_2$F, there are three additional
orbitals resulting from the F-centered $sp^2$-type hybrids ($\sigma_F$).
In contrast to C$_2$H, the $\sigma_{CF}$ and $p_z$ orbitals in C$_2$F are more localized,
implying stronger binding and enhanced stability of the fluorinated system (Table \ref{spreads}).

\begin{figure}[tbp]
\includegraphics[width=0.48\textwidth, angle=0]{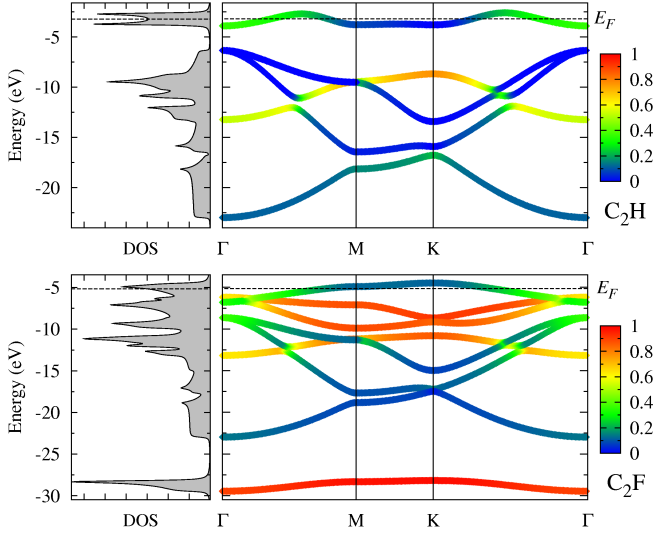}
\caption{(Color online) Band structure and density of states (DOS) for C$_2$H and C$_2$F
obtained from non-spin-polarized calculations. Contributions from the impurity orbitals ($s$ for H and $sp$ for F) are shown by color.}
\label{dosbands}
\end{figure}

In Fig.~\ref{dosbands}, we show the density of states (DOS) and band structure projected onto the orbitals associated with the MLWF
localized at the impurity orbitals of C$_2$H and C$_2$F. In both cases there is a density of states at
the Fermi level, indicating the presence of uncompensated electrons, which are localized mainly at the $p_z$-orbital of
carbon with some contribution from the impurity orbitals.

\begin{figure}[!tbp]
\includegraphics[width=0.47\textwidth, angle=0]{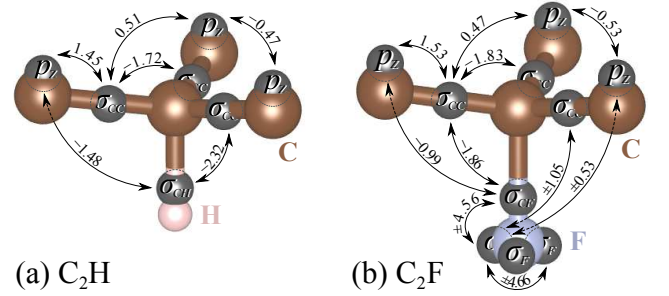}
\caption{(Color online) The most relevant hopping parameters (in eV) between different Wannier orbitals for C$_2$H (a) and C$_2$F (b).
The orbitals are shown schematically by gray labelled spheres, localized at the corresponding MLWF centers.
Positive and negative values correspond to destructive and constructive overlap between the involved orbitals, respectively.}
\label{hoppings}
\end{figure}

The MLWF formalism allows us to represent the electronic Hamiltonian in the tight-binding form,
\begin{equation}
H=\sum_{i} \varepsilon_i n_i + \sum_{i\neq j} t_{ij} c_i^{\dag}c_j,
\label{wan_hamilt}
\end{equation}
where summation runs over the Wannier orbitals, $\varepsilon_i$ is the energy of the $i$th orbital, $t_{ij}$ is the hopping parameter between $i$th and $j$th MLWFs, and $c_i^{\dag}$ ($c_j$) is the creation
(annihilation) operator of electrons localized at the $i$th ($j$th) MLWF. In Fig.~\ref{hoppings}, we show the most relevant hopping parameters
obtained from the non-spin-polarized Hamiltonian for C$_2$H and C$_2$F. In contrast to graphene, where nearest-neighbor hoppings only are essential
to describe the main characteristics of the ground-state electronic spectrum, the case of graphene derivatives is considerably more complicated due to the lowering of symmetry.
In view of the complexity of the Hamiltonian, a simple analytical model, which would be appropriate for studying magnetic
properties of C$_2$H and C$_2$F, is not readily available.

If the spin polarization is switched on, the exchange interactions split the energies of
electrons with opposite spins, resulting in the formation of magnetic states. Interestingly, spin-polarized DF calculations yield different ground states for C$_2$H and C$_2$F, corresponding to the ferromagnetic (FM) and
antiferromagnetic (AFM) configurations with a small band gap ($\sim$0.5--1.0 eV), respectively \cite{Zhou,Boukhvalov2010,L-Li}.
Exploiting exponential localization of the MLWF in real space \cite{Brouder}, the magnetic moments can be associated with the respective Wannier orbitals.
Accordingly, the exchange splitting on the $i$th Wannier orbital can be calculated from
diagonal elements of the spin-polarized MLWF Hamiltonian as $\Delta_i = H_{ii}^{\uparrow}-H_{ii}^{\downarrow}$, whereas the
corresponding magnetic moment is easily obtained from the density of states (DOS) $g^{\sigma}_i(\varepsilon)$ projected onto the $i$th MLWF,
$M_i=\int_{-\infty}^{E_F} d\varepsilon \left[ g^{\uparrow}_i(\varepsilon) - g^{\uparrow}_i(\varepsilon) \right]$. The corresponding values are listed in Table \ref{spreads}
for different Wannier orbitals. Apart from the magnetic moment localized in the $p_z$ orbital, there is a non-negligible contribution from the impurity orbitals,
which is more pronounced for C$_2$H ($\sigma_{CH}$ orbital). Below, we show that this contribution plays an essential role in determining the magnetic
behavior of C$_2$H.

    \begin{table}[!bt]
    \centering
    \caption[Bset]{Linear spreads ($\Omega$), magnetic moments ($M$) and exchange splittings ($\Delta$) associated with the Wannier orbitals of C$_2$H and C$_2$F.
The orbitals are denoted according to Fig.~\ref{wfs}. The spreads are given for the non-spin-polarized orbitals.}
    \label{spreads}
 \begin{tabular}{cccccccccc}
      \hline
      \hline
 &  \multicolumn{3}{c}{C$_2$H} & & \multicolumn{4}{c}{C$_2$F} \\
\cline{2-4}
\cline{6-9}
               & \ $\sigma_{CC}$ \ & \ $\sigma_{CH}$ \ & \ $p_z$ \   &      & \ $\sigma_{CC}$ \ & \ $\sigma_{CF}$ \ & \ $p_z$ \ & \ $\sigma_{F}$ \ \\
     \hline
$\Omega$, \AA  &  \    0.86  \   &  \    1.02  \   &  \   1.25   \      &      &   \  0.84  \     & \      0.79   \  & \  1.11 \ & \    0.71  \   \\
$|\Delta|$, eV &  \,    0.29  \,   &  \,    0.62  \,   &  \,   2.54   \,      &      &   \,  0.24  \,     & \,      0.41   \,  & \,  1.92 \, & \,    0.24  \,   \\
$M$, $\mu_B$   &  \    0.02  \   &  \    0.17  \   &  \   0.76   \      &      &   \  0.01  \     & \      0.05   \  & \  0.59 \ & \    0.02  \   \\
      \hline
      \hline
    \end{tabular}
    \end{table}

Owing to the possibility of a localized representation of magnetic moments in C$_2$H and C$_2$F, the magnetic interactions can be analyzed in terms of the Heisenberg model,
\begin{equation}
H=-\sum_{i \ne j}J_{ij}{\bf s}_i \cdot {\bf s}_j,
\label{heff}
\end{equation}
where $J_{ij}$ is the exchange coupling between $i$ and $j$ sites, and ${\bf s}_i$, ${\bf s}_j$ are unit vectors pointing in the
direction of local magnetic moments at sites $i$ and $j$, respectively. Thus, we consider the mapping of DF calculations onto {\it classical} Heisenberg model \cite{Lichtenstein}.
At this mapping, the exchange parameters can be calculated
via the second variations of the total energy with respect to infinitesimal rotations of the magnetic moments.
For a lattice with a basis, this can be written in the following form \cite{Lichtenstein}
\begin{equation}
J_{ij}^{\alpha \beta}=\frac{1}{4\pi} \int_{-\infty}^{E_F} d\varepsilon ~ \mathrm{Im} \left[ \Delta_{\alpha} G_{ij}^{\alpha \beta \downarrow}(\varepsilon)\Delta_{\beta}G_{ji}^{\beta \alpha \uparrow}(\varepsilon) \right],
\label{Jij}
\end{equation}
where $i,j$ and $\alpha,\beta$ denote unit cell and sublattice (orbital) indices, respectively. $\Delta_{\alpha(\beta)}$ is the exchange splitting of the orbital $\alpha$
($\beta$), $E_F$ is the Fermi level, and $G_{ij}^{\alpha \beta \sigma}(\varepsilon)$
is the real-space Green's function, whose reciprocal-space representation is given by,
\begin {equation}
G_{\bf k}^{\sigma}(\varepsilon)=\left[ \varepsilon - H^{\sigma}({\bf k}) + i\eta \right]^{-1},
\end{equation}
$\eta \rightarrow +0$. Here, $H^{\sigma}({\bf k})$ is the reciprocal-space Hamiltonian for spin $\sigma=\uparrow,\downarrow$, whose matrix elements are obtained in the MLWF basis
from spin-polarized DF calculations, corresponding to the most favorable magnetic state (FM for C$_2$H, and row-wise AFM for C$_2$F).
The Green's functions are evaluated by taking $\approx$10$^6$ {\bf k}-points in the full Brillouin zone and the parameter $\eta=0.02$.

\begin{figure}[!tbp]
\includegraphics[width=0.49\textwidth, angle=0]{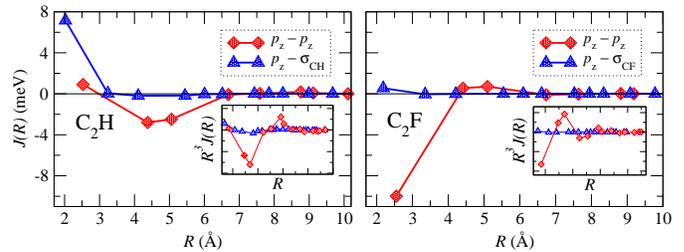}
\caption{(Color online) Calculated exchange parameters as functions of the distance between the $p_z$--$p_z$ (red curve) and $p_z$--$\sigma_{CH(CF)}$ (blue curve) orbitals in C$_2$H (left panel)
and C$_2$F (right panel). The interactions between the other orbitals are negligibly small and not shown.
In the inset, the $J(R)$ parameters are shown with a prefactor $R^3$.}
\label{exchange}
\end{figure}

In Fig.~\ref{exchange}, we show the calculated exchange parameters for both C$_2$H and C$_2$F as functions of the distance between the pairs of the relevant orbitals, i.e., $p_z$--$p_z$ and $p_z$--$\sigma_{CH(CF)}$.
The other orbitals carry negligible moments and thus do not contribute to the exchange. The leading exchange interaction is different in sign for the two systems.
In the case of C$_2$H, it corresponds to FM interactions between the $p_z$ and $\sigma_{CH(CF)}$ orbitals, whereas for C$_2$F, it originates from the $p_z$--$p_z$ AFM coupling.
Compared to C$_2$H, where the leading FM term is followed by a weaker AFM contribution, the interactions in C$_2$F are almost negligible
at larger distances and determined by the first AFM term only. Although for both systems the $p_z$--$p_z$ exchange parameters change their sign with distance, this interaction term decays faster than $R^{-3}$
(see the inset of Fig.~\ref{exchange}). This behavior indicates the absence of RKKY-like interactions, which is not surprising in view of insulating character of both systems.
Importantly, while the magnetic moments arise primarily from the $p_z$ orbitals,
the dominant contribution to the exchange in C$_2$H comes from the FM interaction between the spins localized on the unsaturated carbon ($p_z$) and impurity ($\sigma_{CH}$) orbitals. This contribution is therefore responsible
for the previously reported difference between the ground states of C$_2$H and C$_2$F at the DF level \cite{Zhou,Boukhvalov2010,L-Li}.

In fact, the difference of an order of magnitude between the nearest-neighbor interactions in C$_2$H and C$_2$F can qualitatively be explained
by the difference in the overlap between the $p_z$ and $\sigma_{CH(CF)}$ orbitals, resulting from the unequal spreads (Table \ref{spreads}) and
interorbital distances (Fig.~\ref{wfs}). Thus, this kind of exchange interaction is directly related to the overlap of the wave functions involved.
The second contribution to the exchange ($p_z$--$p_z$ interactions)
exhibits a more complicated behavior, which is typical for the indirect superexchange interactions.
By comparing the hopping parameters given in Fig.~\ref{hoppings}, it can be deduced that the difference in the $p_z$--$p_z$ coupling between C$_2$H
and C$_2$F is related to the presence of the $2p$ fluorine orbitals ($\sigma_F$) in C$_2$F, which provide an additional superexchange pathway compared
to C$_2$H. For both systems, however, a conventional perturbative treatment of the superexchange (e.g., in terms of the relation $J_{ij}\sim t^2_{ij}/U$)
is probably not correct due to the small magnitude of the Coulomb interactions ($U$) in $sp$ systems.

\begin{figure}[!tbp]
\includegraphics[width=0.40\textwidth, angle=0]{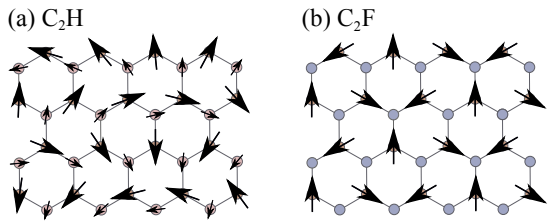}
\caption{(Color online) Spin structures corresponding to a classical ground state for: (a) C$_2$H (incommensurate spin spiral), and (b) C$_2$F (120$^{\circ}$ N\'{e}el state).}
\label{spiral}
\end{figure}

We now turn to the problem of magnetic ordering in C$_2$H and C$_2$F. In both cases, the presence of a significant AFM interaction on the triangular lattice of $p_z$ orbitals suggests the instability of any collinear
spin alignment \cite{Wannier1950}. Surprisingly, the exchange interactions in C$_2$F represent almost an ideal realization of the Heisenberg model with nearest-neighbor AFM couplings ($J=-10$ meV),
whose classical solution corresponds to the 120$^{\circ}$ N\'{e}el state [Fig.~\ref{spiral}(b)].
In contrast to C$_2$F, the ground state of C$_2$H is not obvious due to the presence of a larger number of competing interactions of different signs.
In this case, we seek the ground state spin configuration in the general form
of a coplanar spin spiral,
${\bf S}_{\alpha(\beta)}({\bf r})=S_{\alpha(\beta)}\left(\mathrm{cos}({\bf Q}\cdot{\bf r}+\phi_{\alpha({\beta})}),\mathrm{sin}({\bf Q}\cdot{\bf r} + \phi_{\alpha(\beta)})\right)$, where ${\bf Q}$ is the spiral ordering vector,
and $\phi=\phi_{\alpha}-\phi_{\beta}$ is the phase shift between the spins on different sublattices. Minimizing the total energy with respect to ${\bf Q}$ and $\phi$ we find that the ground state
corresponds to an incommensurate spin spiral with ${\bf Q}$=(0.31,0.79) and $\phi$=41.5$^\circ$. The corresponding spin configuration is shown schematically in Fig.~\ref{spiral}(a). 

The problem of magnetic frustration in C$_2$H and C$_2$F becomes more complicated if the quantum effects are taken into account. If we assume that the unsaturated carbon atoms in C$_2$F carry $S=1/2$ spin,
its ground state would correspond to the three-sublattice magnetic order, similar to the classical case, which is a widely accepted scenario for the spin-1/2 quantum Heisenberg antiferromagnet on a triangular
lattice \cite{three-subl_magn}. However, the magnitude of the local magnetic moment in C$_2$F ($M_{p_z}\approx0.59\mu_B$) strongly suggest itinerant-electron magnetic behavior, which cannot be properly described by the
quantum Heisenberg model. Although multispin models can provide a consistent description of the magnetism on the triangular lattice \cite{Mila}, the way of the parametrization of such models is far from being clear.
The most natural model for studying itinerant-electron magnetism is the Hubbard model. Previous applications of this model to the triangular lattice reveal a spin liquid state at intermediate values of
the $U/t$ parameter with the transition to the ordered N\'{e}el state at higher $U/t$ values \cite{Mila,Spin_liquid}. Given a relatively small magnitude of $U$ in $sp$ materials due to the delocalization of electrons \cite{Ustar},
the spin liquid state appears to be a possible scenario for the ground state of C$_2$F.

The case of C$_2$H turns out to be even more complicated due to the presence of the two magnetic sublattices comprising the honeycomb lattice. Although recent studies of the Hubbard model on the honeycomb lattice
do not exclude the existence of a spin liquid phase in some range of the $U/t$ parameters \cite{SL_hexagon}, their results cannot be directly applied to C$_2$H having two nonequivalent
sublattices with $M_{p_z}\approx0.76\mu_B$ and $M_{\sigma_{CH}}\approx0.17\mu_B$.
An adequate description of such a system requires a more sophisticated analysis in terms of the multiorbital Hubbard model and represents an important issue for further studies.

In conclusion, we have studied the exchange interactions in C$_2$H and C$_2$F and analyzed the possibilities of the magnetic ordering in these systems. In each case, we have found the presence of
AFM interactions on the triangular lattice of magnetic moments, leading to the instability of the collinear magnetic ordering due to frustration. Although the true quantum ground state remains to be determined, both systems
are expected to exhibit highly correlated magnetic behavior at low temperatures and can be considered as promising candidates for the quantum spin liquid.

We acknowledge support by the EC under the Graphene Flagship (contract no. CNECT-ICT-604391) and by the scientific program no. 14.А18.21.0076 (Russia).

\end{document}